# Superconductivity in graphite-diamond hybrid

Yanfeng Ge,[1,2,†] Kun Luo,[1,2,†] Yong Liu,[1,2] Guochun Yang,[1,2,*] Wentao Hu[1], Baozhong Li[1], Guoying Gao[1], Xiang-Feng Zhou[1], Bo Xu[1], Zhisheng Zhao,[1,*] and Yongjun Tian[1]

[1] *Center for High Pressure Science (CHiPS), State Key Laboratory of Metastable Materials Science and Technology, Yanshan University, Qinhuangdao 066004, China*
[2] *Key Laboratory for Microstructural Material Physics of Hebei Province, School of Science, Yanshan University, Qinhuangdao, 066004, China*

† Y. G. and K. L. contributed equally.
*Corresponding authors: yanggc468@nenu.edu.cn; zzhao@ysu.edu.cn.

Search for new high-temperature superconductors and insight into their superconducting mechanism are of fundamental importance in condensed matter physics. The discovery of near-room temperature superconductivity at more than a million atmospheres ushers in a new era for superconductors. However, the critical task of identifying materials with comparable superconductivity at near or ambient pressure remains. Carbon materials can always lead to intriguing surprises due to their structural diversity and electronic adjustability. Insulating diamond upon doping or external stimuli has achieved superconducting state. Thus, it still has a great opportunity to find superconducting ones with higher transition temperature ($T_c$). Here, we report an intrinsic superconducting graphite-diamond hybrid through first-principles calculations, whose atomic-resolution structural characteristics have been experimentally determined recently. The predicted $T_c$ is approximated at 39 K at ambient pressure, and strain energizing can further boost $T_c$ to 42 K. The strong electron-phonon coupling associated with the out-of-plane vibration of carbon atoms at the junction plays a dominant role in the superconducting transition. Our work demonstrates the great potential of such carbon materials as high-$T_c$ superconductors, which will definitely attract extensive research.

One of the urgent and challenging tasks in condensed matter physics is preparing high-temperature superconductors and exploring their superconducting mechanism. Recently, a conventional superconductor represented by hydrides (203 K for $H_3S$[1], 250 K for $LaH_{10}$[2], and 262 K for $YH_9$[3]), whose superconducting transition temperatures are much higher than the record of nonconventional ones (164 K for compressed cuprate material[4]), brings the notion of room-temperature superconductivity within reach. Theoretical predictions guide these pioneering discoveries based on the Bardeen-Cooper-Schrieffer (BCS) and Migdal-Eliashberg theories[5-9], which offer direct insight into the design of high-$T_c$ conventional superconductors. However, more than one million atmospheres are required to stabilize these hydrides or turn them into superconducting states[10-12]. Thus, finding stable or metastable superconducting materials at lower or ambient pressure is necessary for practical use.

Significant progress has been made in achieving the superconductivity of carbon materials upon doping, strain, and compression. Recent studies mainly focus on the four kinds of carbon forms (e.g., diamond, graphite, graphene, and amorphous carbon). Boron-doped[13] and compression-shear deformed diamonds[14] exhibit superconductivity with $T_c$ values of 4 K and 12.4 K, respectively. Furthermore, the latter provides a new routine for driving materials into superconducting states via strain engineering[14]. $CaC_6$, an intercalated graphite compound, has a $T_c$ value of 11.5 K[15]. Magic-angle bilayer graphene demonstrates unconventional superconductivity ($T_c$=1.7 K) and many novel physical phenomena[16,17]. B-doped amorphous quenched carbon (Q-carbon) shows $T_c$ values of 36 K and 55 K at 17% and 27% boron concentration, respectively[18,19]. Unfortunately, the inherent structural uncertainty of Q-carbon masks the superconductivities origin. Overall, these pioneering studies have shown that carbon materials have the potential to become high-$T_c$ superconductors.

Carbon can form a plethora of allotropes with distinct physical properties due to its versatility in bonding mode of sp, $sp^2$, and $sp^3$ hybridization; thus, a great deal of studies have been done to explore new carbon forms with exotic properties[20-24]. Recently, a kind of diamond-graphite hybrid structures, composed of graphitic layers inserted within the Pandey (2×1) reconstructed surfaces of diamond {113} planes, were proposed through the observation of natural impact diamonds by high resolution transmission electron microscopy (HRTEM)[20,21]. It is predicted that the hypothetical crystal structures of this kind of hybrid carbon would possess unique mechanical, electronic, and thermal properties[20]. Meanwhile, we identified the coherent graphite-diamond interfaces comprised of four structural motifs in partially transformed graphite recovered from static compression, by using atomic-resolution high-angle annular dark-field scanning transmission electron microscope (HAADF-STEM)[25]. There are four representative hypothetical crystal structures of graphite-diamond hybrid, *i.e.* Gradia-CO, Gradia-CA, Gradia-HB, and Gradia-HC[25]. These hybrid structures are expected to combine the advantage of graphite/graphene and diamond, exhibiting the extraordinary mechanical properties (i.e. ultra-high hardness and toughness) and adjustable electrical properties, which may be engineered to produce new generation family of carbon materials[24,25].

The motivation for investigating the superconductivity of Gradia-CO crystal structure in this work stems from the emergence of metallicity mainly contributed by carbon atoms at the junction[25]. In contrast with the reported superconductivity in various carbon forms induced by doping and strain[13-19], Gradia-CO shows intrinsic superconductivity with $T_c$ up to 39 K. The unprecedented superconductivity induced by carbon atoms at graphite-diamond interface demonstrates the great potential for developing new superconducting electronic devices based on graphite-diamond hybrid materials.

We construct the crystal structures of Gradia-CO on the basis of the atomically resolved interface structures of graphite-diamond hybrid determined by HAADF-STEM[25], as shown in Fig. 1(a). The structures, termed Gradia-CO($m,n$)[25], consist of variable diamond and graphite units. The $m$ and $n$ are the numbers of hexagonal units formed between adjacent graphene-like layers (indicated by red atoms) in the confined graphite and diamond domains, respectively. The coherent junction is (110)$_{CD}$, and the two domains have the relation [010]$_G$ // [1$\bar{1}$0]$_{CD}$. Here we systematically study the physical properties related to the superconductivity of Gradia-CO($m,n$) based on first-principles calculations.

Considering the structural similarity of Gradia-CO($m,n$), we first focus on the discussion of Gradia-CO(7,5). After high-precision structural optimization at ambient pressure, Gradia-CO(7,5) stabilizes into a monoclinic structure (space group P2$_1$/m). The computed lattice constants are $a_0$=18.54 Å, $b_0$=2.48 Å, and $c_0$=3.54 Å with non-perpendicular $a$ and $c$ axes (Fig. 1a). The graphite part has a reduced interlayer spacing of *ca.* 3.1 Å, due to the constraint of the junction, which is consistent with the interlayer spacing of previously reported "compressed graphite" [26-28]. The C-C bond length (1.64 Å) at the junction becomes a little longer with respect to that of graphite (1.42 Å) and diamond (1.55 Å). However, their covalent feature is well maintained given the obvious bonding electron pairs between the two adjacent carbon atoms (Fig. 1b).

The structural stability of the material is a prerequisite for practical application. The absence of imaginary frequency of Gradia-CO(7,5) in the whole Brillouin zone confirms its dynamic stability (Fig. 1d). The highest phonon vibration frequency is about 1550 cm$^{-1}$, comparable to graphite intercalated compounds and metal-decorated graphene[29,30]. First-principles molecular dynamics (MD) simulations for Gradia-CO(7,5) were carried out for 4 ps at 300 K (Fig. 1c) to verify its temperature-dependent stability. Its total energy has a stable range of oscillation, and neither bond breakage nor significant structural deformation is observed, demonstrating that Gradia-CO(7,5) is thermally stable at room temperature.

The electronic band structure and projected density of states (PDOS) of Gradia-CO(7,5) (Fig. 2a) show inherent metallicity because of several electron bands across the Fermi level. More intriguingly, its metallicity mainly comes from the junction's two types of C atoms. One is the $C_1$ atom that coordinates with three neighboring carbon atoms, showing an unsaturated bonding feature in the diamond part (Fig. 2b). The other is the $C_2$ atom forming a zigzag edge in the graphite part. Moreover, the contribution of the $p_x$ and $p_z$ orbitals to metallicity is dominant (Fig. S1). However, there appear to be relatively flat dispersion relations than other energy ranges, leading to the high DOS around the Fermi level (Fig. 2a). These features favor the strong electron–phonon coupling to induce superconductivity.

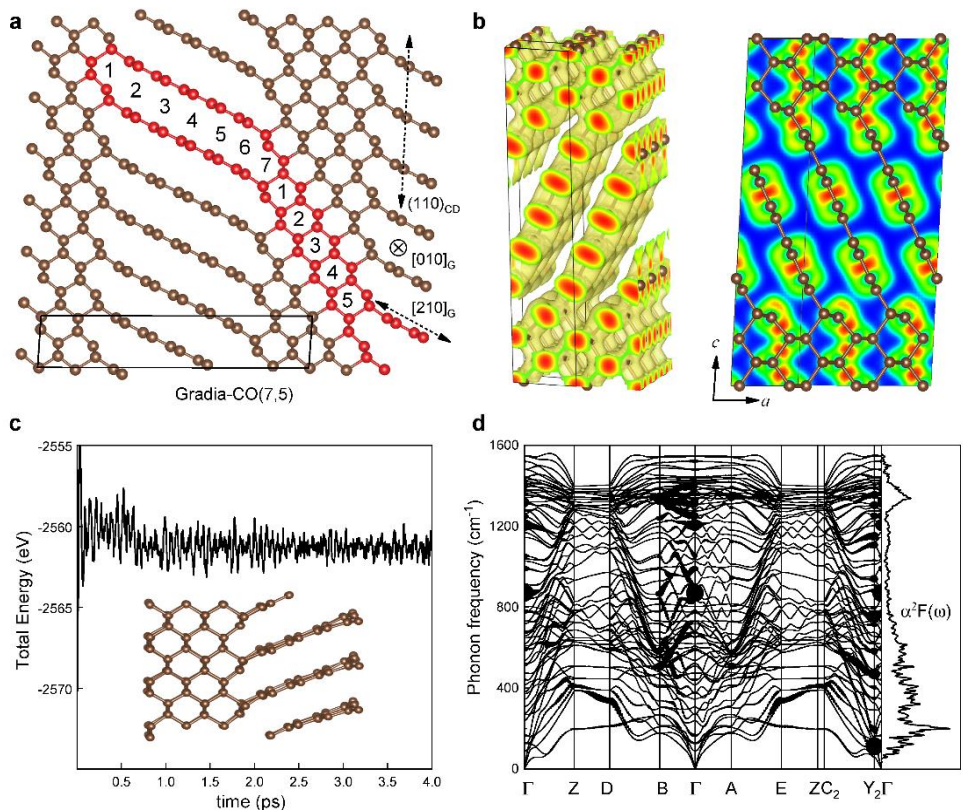

**Fig. 1| The structure and stability of Gradia-CO(7,5)**. **a**, Nomenclature of Gradia-CO(*m,n*). *m* and *n* are the numbers of hexagonal units between adjacent graphene-like layers in the confined graphite and diamond domains, respectively. The schematic is the structure of Gradia-CO(7,5). **b**, Electron localization function (ELF) of Gradia-CO(7,5). The red (blue) indicates the maximum (minimum) value of 1 (0) in the two-dimensional slice. **c**, The molecular dynamics simulation of Gradia-CO(7,5) at 300 K. **d**, Phonon dispersion and Eliashberg function $\alpha^2F(\omega)$ of Gradia-CO(7,5). The size of the black circles is proportional to the phonon linewidths.

We then performed electron-phonon coupling (EPC) calculations of Gradia-CO(7,5) to explore its potential superconductivity by using the density functional perturbation theory (DFPT). The plotted phonon linewidths can indicate the frequency-dependent EPC (Fig. 1d). The vibration modes covering the entire frequency range of Gradia-CO(7,5) contribute to EPC, but their relative contributions are distinct. Specifically, the contribution of high-frequency phonons around 1400 cm$^{-1}$ to EPC is relatively small. In contrast, the low-frequency phonons are the main EPC participant. The phonons between 200 and 400 cm$^{-1}$ have the largest phonon linewidths, making the major contribution of Eliashberg function a$^2$F ($\omega$). The analysis of the atomic vibration mode confirms that the phonons strongly coupled with electrons are the out-of-plane atomic vibration of the C layer, as shown by the black arrows in Fig. 2b. The resultant electron-phonon coupling constant ($\lambda$) is ~0.90, comparable to $MgB_2$ and $CaC_6$[31-33]. However, low-frequency phonons with strong coupling cause a not-so-high logarithmically averaged phonon frequency $\omega_{log}$ = 381 K between $MgB_2$ and $CaC_6$[31-33]. Employing the effective constant $\mu^*$ of 0.1, the estimated $T_c$ of Gradia-CO(7,5) is about 22 K, higher than $T_c$ values in many carbon-based materials[13,14,34].

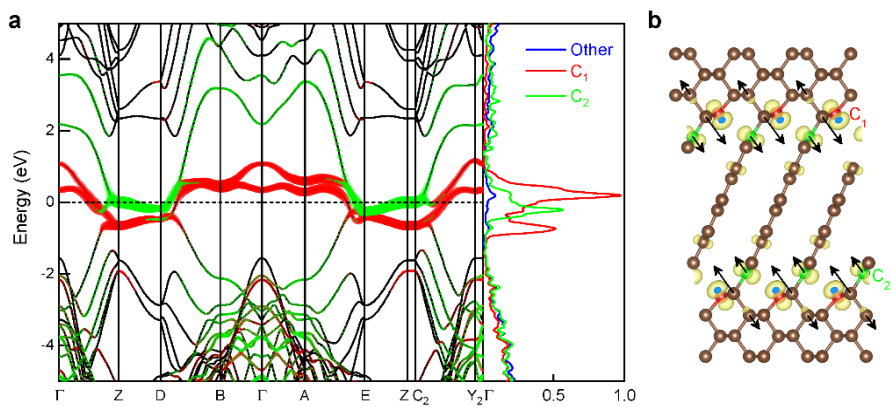

**Fig. 2| The electronic structure of Gradia-CO(7,5)**. **a**, Electronic band structure with the projected orbitals of $C_1$ and $C_2$ atoms. **b**, Unsaturated-bonding $C_1$ atoms of the diamond part and $C_2$ atoms at the zigzag edge of the graphite part. The electron density isosurfaces show the partial charge density around Fermi level [−1.0, +1.0], and blue arrows indicate the atomic vibration of the low-frequency phonon modes strongly coupled with electrons.

The strain will change the electronic or phonon structure of the material, just like the superconductivity of compression-shear deformed diamond[14]. Hence the present work further explores the effects of uniaxial tensile and compression strain along the *b* axis, which is introduced by adjusting the lattice constant *b* with the strain capacity $\varepsilon = (b − b_0)/b_0 \times 100\%$ and reoptimizing the other two axes. As shown in Fig. 3, under the uniaxial strains ($\varepsilon = \pm3\%$), the electronic structure around the Fermi level of Gradia-CO(7,5) is almost unchanged, except for the slight shift of band at the $\Gamma$ point. That is, the feature of high DOS at the Fermi level is still maintained. However, the strain effect becomes more obviously reflected in the phonon spectra (Fig. S3). Under compression (tensile) strain, the shorter (longer) bond length and stronger (weaker) bond energy can increase (decrease) the frequency of the high-frequency phonon. More importantly, phonon softening occurs in the low-frequency phonons strongly coupled with electrons under compression strain. Upon compression, the peak of $\alpha^2F$ ($\omega$) in the low-frequency region increases, accompanied by the redshift of the peak position (Fig. 3b). As a result, the $\lambda$ is enhanced substantially (Fig. 3c), favoring BCS superconductivity. However, the redshifted peak of $\alpha^2F$ ($\omega$) also brings about the decrease of $\omega_{log}$. To be noted, the opposite phenomenon appears in tensile strain, such as the blue shifted lower peak of $\alpha^2F$ ($\omega$) and the weak EPC (Fig. 3c and Fig. S3). Under compression strain ($\varepsilon = -3\%$), the critical temperature of Gradia-CO(7,5) can reach 40 K.

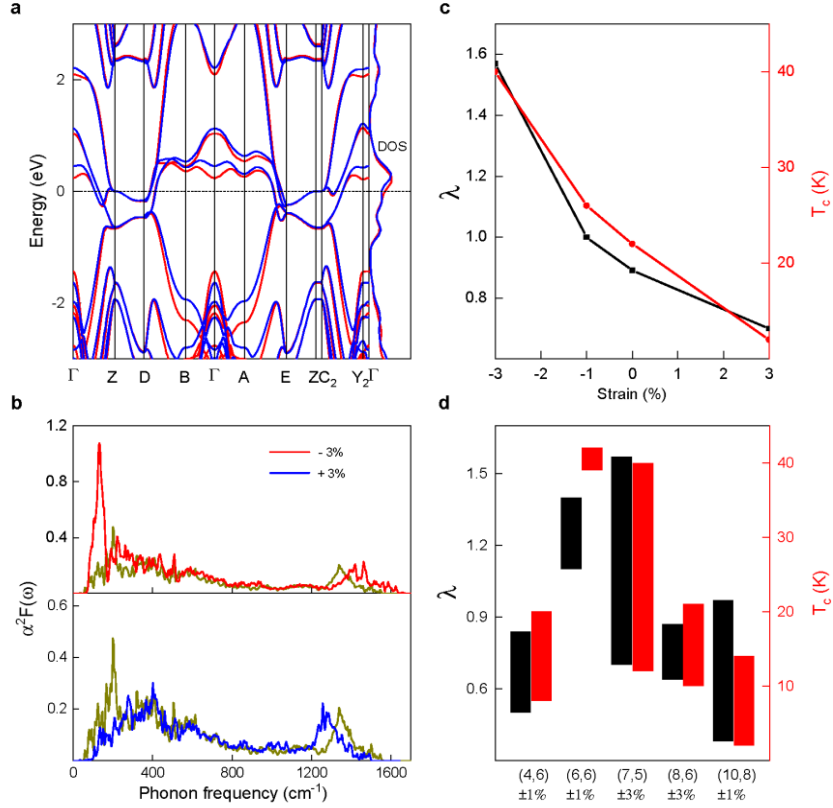

**Fig. 3| Strain and size effects. a**, Electronic band structure and DOS, as well as **b**, Eliashberg function $\alpha^2F(\omega)$ of Gradia-CO(7,5) under different types of strain ($\varepsilon=\pm3\%$). Red (blue) lines are under the uniaxial compressive (tensile) strain along the $b$ axis. To facilitate comparison, the dark yellow lines in **b** are the results of the strain-free structure. **c**, $\lambda$ and $T_c$ of Gradia-CO(7,5) within strains from +3% to −3%. (d) $\lambda$, and $T_c$ of Gradia-CO($m,n$) within their dynamically stable strain range.

Finally, we explore the size effect on the superconductivity, such as the different widths of graphite or diamond in Gradia-CO($m,n$). The five considered structures have strong EPC to induce superconductivity (Fig. 3d). However, their DOS values at the Fermi level closely correlate with the size: the values decrease with the increase of size, and Gradia-CO(4,6) has the highest value of 0.31 states/eV/atom. However, the larger size reduces the strongly coupled phonon frequency by around 200 cm$^{-1}$ (Figs. S4 and S5). As a result, the variable behaviors of electronic structures and phonon spectra with the size have the opposite effects on $\lambda$. The highest-$T_c$ superconductivity appears in Gradia-CO(6,6) with a computed $T_c$ of 39 K, mainly originating that its $\lambda$ of 1.35 is much higher than 0.9 in Gradia-CO(7,5). Finally, we also explore the strain effect on all structures. Here, the stress range of each structure depends on the allowable stress for dynamic stability. Combining all the conditions, the highest $T_c$ of Gradia-CO(6,6) is 42 K.

In summary, we have studied the phonon-mediated superconductivity of Gradia-CO($m,n$), a new kind of graphite-diamond hybrid material. The material's electronic states around the Fermi level are mainly from the p$_x$ and p$_z$ orbitals and distribute near the diamond–graphite junction. More intriguingly, their superconductivity also originates from the out-of-plane vibrations of carbon atoms around the junction. This material shows strong EPC with a high $T_c$ value of 39 K, superior to most reported carbon materials and elemental superconductors. Still, their $T_c$ values can be further elevated through strain engineering. The current finding not only predicts a new type of superconducting carbon material—graphite-diamond hybrid, but also demonstrates the feasibility of hybrid interface superconductivity in carbon materials, which will greatly promote the extensive theoretical and experimental exploration of high-$T_c$ carbon-based superconductors.

**Methods**

In the calculation of electron-phonon coupling[37], the electron-phonon matrix elements $g_{mn,\nu}(\mathbf{k},\mathbf{q})$ are expressed as

$$g_{mn,\nu}(\mathbf{k},\mathbf{q}) = \left(\frac{\hbar}{2\omega_{\mathbf{q}\nu}m_0}\right)^{1/2} \langle\psi_{m\mathbf{k}+\mathbf{q}}|\partial_{\mathbf{q}\nu}V|\psi_{n\mathbf{k}}\rangle$$

The electron-phonon coupling strength associated with a specific phonon mode and wavevector $\lambda_{\mathbf{q}\nu}$ is given by

$$\lambda_{\mathbf{q}\nu} = \frac{1}{\omega_{\mathbf{q}\nu}N_F}\sum_{mn,\mathbf{k}}|g_{mn,\nu}(\mathbf{k},\mathbf{q})|^2 \delta(\epsilon_{n\mathbf{k}})\delta(\epsilon_{m\mathbf{k}+\mathbf{q}})$$

The dimensionless parameter electron-phonon coupling constant $\lambda$ is calculated as the Brillouin-zone average of $\lambda_{\mathbf{q}\nu}$:

$$\lambda = \sum_{\mathbf{q}\nu}\lambda_{\mathbf{q}\nu}$$

The Eliashberg spectral function $\alpha^2 F(\omega)$ can be calculated in terms of $\lambda_{\mathbf{q}\nu}$ and the phonon frequencies using:

$$\alpha^2 F(\omega) = \frac{1}{2}\sum_{\mathbf{q}\nu}\omega_{\mathbf{q}\nu}\lambda_{\mathbf{q}\nu}\delta(\omega-\omega_{\mathbf{q}\nu})$$

The Tc was estimated by the Allen-Dynes-modified McMillan formula

$$T_c = \frac{\omega_{\log}}{1.20} \exp\left[-\frac{1.04(1+\lambda)}{\lambda - \mu^*(1+0.62\lambda)}\right]$$

The first-principles calculations were carried out with QUANTUM ESPRESSO[38], using the norm-conserving pseudopotentials[39]. The exchange-correlation effect was in the Perdew–Burke–Ernzerhof (PBE) form[40]. For all the calculations, the kinetic energy cutoff of 40 Ry and Monkhorst–Pack k-mesh of 8 × 32 × 24 was used. The phonon spectrum and electron–phonon coupling were calculated on a 4 × 16 × 12 q-grid by using DFPT[41].

## Data availability
All data that support the findings of this paper are available from the corresponding authors upon request.

## Code availability
Source codes for the first-principles calculations are available from the corresponding authors upon request.

## Acknowledgements

This work is supported by National Natural Science Foundation of China (Nos.52090020, 11904312, 21873017, 91963203, 51772260, U20A20238, 52025026, 52073245, 51722209, 51525205), the National Key R&D Program of China (2018YFA0703400 and 2018YFA0305900), the Natural Science Foundation for Distinguished Young Scholars of Hebei Province of China (E2018203349), and the Talent research project in Hebei Province (2020HBQZYC003).

## Author contributions

G.Y., Z.Z., and Y.T. conceived the project. Y.G., and K.L. implemented the first-principles calculations, Y.G., K. L., Y.L., G.Y., W.H., B.L., G.G., X.Z., B.X., Z.Z., and Y.T. analyzed the data and discussed the results. Y.G., K.L., G.Y., and Z.Z. contributed to the writing of the manuscript.

**Competing interests**
The authors declare no competing interests.